# A Markov Chain Monte-Carlo Approach to Dose-Response Optimization Using Probabilistic Programming (RStan)

Dorsa Mohammadi Arezooji
Department of Electrical Engineering and Computer Science (EECS)
Queen Mary University of London
*London, UK*
d.mohammadiarezooji@se19.qmul.ac.uk

*Abstract*: A hierarchical logistic regression Bayesian model is proposed and implemented in R to model the probability of patient improvement corresponding to any given dosage of a certain drug. *RStan* is used to obtain samples from the posterior distributions via Markov Chain Monte-Carlo (MCMC) sampling. The effects of selecting different families of prior distributions are examined and finally, the posterior distributions are compared across *RStan*, and two other environments, namely *PyMC*, and *AgenaRisk*.

*Keywords*: Markov Chain Monte Carlo (MCMC), Hamiltonian Monte-Carlo (HMC), Bayesian Statistics, Stan, Probabilistic Programming, Hierarchical Logistic Regression

## 1. Introduction

In recent years, Bayesian inference has revolutionized many industries including drug development. A vital step in developing any drug is conducting extensive clinical trials which has several drawbacks. First, it involves testing the drug on a large number of subjects which is costly and possibly time-consuming; and second, some subjects would inevitably receive suboptimal treatment (Berry, 2006). Furthermore, the target group for some drugs might not be large enough to allow conducting a statistically reliable clinical trial. Hence, the purely "frequentist" approach might not be a proper choice in these situations (Yin, 2013). Additionally identifying the optimal dosage of a drug is crucial, as a low dosage might not have the desired effect and a high dosage could potentially cause toxicity and endanger the patient's life (Abraham and Daurès, 2004). To avoid the mentioned drawbacks, a number of Bayesian statistical models have been developed to determine patient-specific optimal dosage (Whitehead and Brunier, 1995), or to simulate the effects of selecting different dosages in dose-response studies (Kwon *et al.*, 2016).

### 1.1. The Bayesian Approach

Bayesian models are great candidates for clinical trials since they continuously learn from the data (and other nodes in a hierarchical structure) to improve estimates or predictions. The Bayesian approach incorporates prior information and belief from different sources into the model. Instead of discretely updating the knowledge in large steps, each one corresponding to a separate trial or trial phase, the Bayesian approach gradually updates the posterior probabilities via Bayes' theorem (Berry, 1993). In case of a Bayesian model with parameters $\theta$, and data $x$, the posterior can be evaluated using Bayes' theorem:

$$p(\theta|x) = \frac{p(x|\theta)p(\theta)}{p(x)} = \frac{p(x|\theta)p(\theta)}{\int_\theta p(x|\theta)p(\theta)\,\mathrm{d}\theta}$$

In the case that the prior $p(\theta)$ and the likelihood $p(x|\theta)$ are conjugates, the posterior $p(\theta|x)$ will conveniently have a closed-form solution (Raiffa and Schlaifer, 1961); otherwise, the integral is evaluated using numerical methods. These numerical methods can be categorized into two categories: first, methods using variants of the Laplace approximation to integrate out latent variables (Rue, Martino and Chopin, 2009; Kristensen *et al.*, 2016), and second, methods using variants of Markov Chain Monte-Carlo (MCMC) (Robert and Casella, 1999). The first category is vastly used since these methods are computationally fast and can also be implemented by users without expert domain knowledge in computational statistics. Nevertheless, such methods have limited approximation accuracy and could be restricting in regard to the types of models that can be implemented.

### 1.2. Markov Chain Monte-Carlo (MCMC)

MCMC sampling is a powerful tool with applications in different areas such as finance (Herzog and Lord, 2002), cognitive science (Raaijmakers *et al.*, 2015), pharmaceutical drug trials (Chang, 2010), and many more. There are a number of complex high-dimensional cases where MCMC is the only known solution that yields good results in an acceptable amount of time. For instance, Dyer et al. used MCMC sampling to approximate the volume of a convex body in an n-dimensional Euclidean space (Dyer, Frieze and Kannan, 1991). MCMC algorithms aim at extracting information regarding distributions, in particular information about posterior distributions in Bayesian inference. They can be implemented to estimate the parameters of a certain distribution through randomly generating samples from that distribution (Rosenthal, 2009). MCMC performs quite well even in cases when all that is known is how to calculate the probability density for the different samples drawn from the distribution in question (Gilks, Richardson and Spiegelhalter, 1995). There are two main concepts that make up MCMC: the Monte-Carlo method, and Markov Chains.

*1.2.1. Monte-Carlo Methods*

In theory, any problem in a probabilistic framework can be solved using Monte-Carlo methods (van Ravenzwaaij, Cassey and Brown, 2018). Monte-Carlo refers to a class of algorithms which estimate the characteristics of a distribution by evaluating the properties of random samples drawn from that distribution. Monte-Carlo methods are particularly useful when calculating the statistical properties of a large enough collection of samples is much simpler than analytically calculating those properties from the distribution. The case for using Monte-Carlo methods is even stronger when those random samples are easy to generate while the distribution's equations are hard to solve (van Ravenzwaaij, Cassey and Brown, 2018).

*1.2.2. Markov Chains*

The Markov chain property of MCMC states that the random samples drawn from the target distribution are generated by a specific sequential process (Gilks, Richardson and Spiegelhalter, 1995). Each random sample *i* is used to generate the following random sample *i+1*. Although each sample depends on the previous one, it does not depend on any other samples before that (referred to as the "Markov" property).

MCMC methods are especially useful in Bayesian inference due to the focus on posterior distributions which would otherwise be difficult to evaluate through analytical methods. For instance, assume we have a collection of numbers generated from a distribution. Now one way to go about calculating the mean of the distribution, is to create a Markov chain that has the distribution as its equilibrium state and uses Monte-Carlo sampling to generate the random samples that make up the chain. Then, to calculate the mean of the distribution, one could simply calculate the mean of the generated samples. This is MCMC in a nutshell. This approach is backed up by the law of large numbers, stating that integral described by the expected value of a random variable (RV) can be approximately evaluated by calculating the mean of independent identically distributed (i.i.d) samples drawn from that RV.

*1.2.3. Hamiltonian Monte-Carlo (HMC)*

The Hamiltonian Monte-Carlo (HMC) algorithm (Duane *et al.*, 1987; Neal, 2011) is an MCMC method which has seen recent widespread application in Bayesian statistics. This is in part due to the fact that these algorithms can generate close to i.i.d Markov chains by only needing to calculate the gradient of target distribution's log-density. However, direct application of HMC might not work as intended when applied to hierarchical Bayesian models as these models incorporate strong non-linearities and significantly different scaling properties (Betancourt, 2013).

*1.3. Stan*

There are currently a number of statistical software programs available that offer MCMC sampling. These programs allow users to define a Bayesian model and evaluate properties of the posterior distribution. The first convenient implementation of MCMC began with the BUGS (Bayesian inference Using Gibbs Sampling) software package (Spiegelhalter & Gilks, 1992). BUGS allowed users to code their model's equations with a syntax similar to R. It would then compile the model and generate the Markov chains and generate samples using Gibbs sampling or random walk Metropolis (RWM) for each parameter. Although BUGS had a major role in the widespread use of MCMC sampling, it is no longer being supported. However, it leaves behind two successors: OpenBUGS (Lunn *et al.*, 2000), and JAGS (Just Another Gibbs Sampler) (Plummer, 2003). The development and release of "Stan" in 2012 marks the latest advances made in open source MCMC.

Stan (Stan Development Team, 2015), named after Stanislaw Ulam, one of the developers of the Monte-Carlo method (Metropolis and Ulam, 1949), is a probabilistic programming language, created to facilitate statistical modeling, in particular Bayesian inference models. It comes with interfaces in Python, Julia, Matlab, and R. It is also more flexible than BUGS or JAGS, and outperforms them in terms of speed when dealing with complex models (Gelman, Lee and Guo, 2015). Stan, also scales up better as the size of the dataset increases, which is in part because of its advanced algorithms and efficient use of memory resources (Gay, 2005). Multi-level models that have matrix parameters (a set of different parameter values for each group) are somewhat slow in BUGS and JAGS as they are based on Gibbs sampling, which does not incorporate covariance matrices very well (Gelman, Lee and Guo, 2015). In these cases, Stan is the practical choice.

Further, unlike BUGS or JAGS, Stan does not require the user to specify a prior distribution for every parameter as long as the posterior is proper. Instead, it assumes a uniform prior defined over the space of possible values for all parameters whose prior distributions have not been explicitly declared. Also in the case of HMC, Stan does not require the user to specify the derivatives of the target distribution, which is crucial for the HMC algorithm. Another benefit of using Stan is the ability to easily reuse previously created models.

As powerful as Stan is, it has some limitations. First, unlike BUGS and JAGS, it can not be used for models with discrete parameters except for some finite mixture models where those parameters can be averaged over (Gelman, Lee and Guo, 2015). Second, if a subset of modeled data is missing, the Stan model needs to be altered to account for these missing values (Gelman, Lee and Guo, 2015).

In cases where sampling is not as fast as needed, it is useful to take advantage of parallel computing and distribute the computational load between multiple processors. Since there are usually more than one Markov chain running in implementations of MCMC (for increased accuracy) and then being combined for inference, parallelization helps with speeding up the process (Shirley, 2011).

## 2. Background and Method

As discussed earlier, there exists a closed-form solution for cases where the likelihood and the prior are conjugates. In the context of drug trials, in its simplest form, each experiment corresponding to a certain dosage could be regarded as a Bernoulli trial with two outcomes depending on whether the subject's condition improved or not (0 and 1). In other words, with the chances of improvement (denoted by $q$) equal to $x$, there will be $n$ successes out of $N$ for every experiment, which can be modelled by a binomial distribution:

$$P(n,N|q=x) = \binom{N}{n} x^n (1-x)^{N-n}$$

The usual conjugate prior for a binomial distribution is a beta distribution with parameters $\alpha$ and $\beta$:

$$P(q=x) = \frac{x^{\alpha-1}(1-x)^{\beta-1}}{B(\alpha,\beta)}$$

$$B(\alpha,\beta) = \frac{\Gamma(\alpha)\Gamma(\beta)}{\Gamma(\alpha+\beta)}$$

It is noteworthy that $\alpha$ and $\beta$ are the hyperparameters, thus should not be confused with the underlying model's parameter ($q$). Now, incorporating these two probability mass functions into Bayes' theorem would give:

$$P(q=x|n,N) = \frac{P(n,N|x)P(x)}{\int_y P(n,N|y)P(y)dy}$$

$$= \frac{\binom{N}{n}x^{n+\alpha-1}(1-x)^{N-n+\beta-1}/B(\alpha,\beta)}{\int_{y=0}^{y=1}\left(\binom{N}{n}x^{n+\alpha-1}(1-x)^{N-n+\beta-1}/B(\alpha,\beta)\right)dy}$$

$$= \frac{x^{n+\alpha-1}(1-x)^{N-n+\beta-1}}{\int_{y=0}^{y=1}\left(x^{n+\alpha-1}(1-x)^{N-n+\beta-1}\right)dy}$$

$$= \frac{x^{n+\alpha-1}(1-x)^{N-n+\beta-1}}{B(n+\alpha,N-n+\beta)}$$

This shows that the posterior distribution is another beta distribution with parameters $n+\alpha$ and $N-n+\beta$.

Although it is rather straightforward to calculate the posterior probability in the case of conjugacy, a beta prior might not be the best representation of the prior belief about the model. In other cases, numerical sampling methods such as MCMC can provide acceptable results.

*2.1. HMC and Hierarchical Models*

In a hierarchical model, the parameters are organized into conditionally independent groups. For a one level hierarchical model with a set of parameters $\theta$ and hyperparameters $\phi$ and data $y$, that consists of $n$ groups, the posterior is evaluated using Bayes' theorem:

$$p(\theta,\phi|y) \propto \prod_{i=1}^{n} p(y_i|\theta_i)p(\theta_i|\phi)p(\phi)$$

In this hierarchical structure, the hyperparameter, i.e. global parameter, $\phi$ interacts with all $n$ local parameters $\theta$. With each batch of new data $y_i$, the corresponding local parameter $\theta_i$ is updated. This allows the hyperparameter $\phi$ to update all other $n-1$ local parameters once it receives the information from $\theta_i$.

One of the drawbacks of hierarchical models is the fact that since the local parameters are dependent on the global parameters, the structure of the models is highly correlated. This correlation worsens as the number of levels or groups increases (Betancourt and Girolami, 2015). One solution to mitigate the effects of correlation between the layers of hierarchical models is to break their dependencies through parameterization.

Non-centered parameterization (NCP) can be used to replace certain dependencies between the layers with deterministic transformations (Papaspiliopoulos, Roberts and Sköld, 2007). As illustrated in the figure below, NCP translates the indirect dependence between $\phi$ and $y$ through $\theta$ into a direct dependence between $\phi$ and $y$ by parameterizing $\theta$ as $\vartheta$. By doing so, $\phi$ and $\vartheta$ would be independent conditioned on the data.

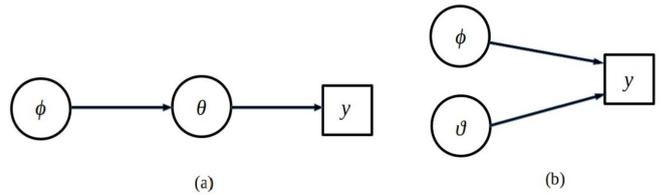

Fig. 1. (a) A one-level hierarchical model without reparameterization, and (b) the same model with NCP

Often in hierarchical models, NCP speeds up the sampling process by several orders of magnitude (Betancourt and Girolami, 2015). This advantage of NCP is more pronounced in HMC compared to other MCMC methods such as RWM. The speed-up in runtime scales with the model's dimensionality to the point that in the most complex problems, HMC is not just the most effective answer, but the only practical one (Betancourt and Girolami, 2015).

*2.2. Data Exploration*

The dataset holds information about a number of drug trial experiments carried out on pigs. Each row includes data corresponding to a certain dosage $d$ of the drug tested on a number of pigs $N$, out of which some $n$ improved after receiving the treatment. There are a total of 71 experimental trials in the dataset.

Table 1. Statistical Summary of the Dataset

|         | Dosage (d) | Total Subjects (N) | Improved Subjects (n) |
|---------|------------|--------------------|-----------------------|
| Min.    | 0.730      | 10                 | 0                     |
| 1st Qu. | 1.185      | 19                 | 1                     |
| Median  | 1.300      | 20                 | 4                     |
| Mean    | 1.276      | 24.49              | 6.183                 |
| 3rd Qu. | 1.380      | 22.50              | 7.500                 |
| Max.    | 1.890      | 52                 | 45                    |

To contrast the frequentist and the Bayesian approach, first the survival ratio $n(d)/N(d)$, is plotted against the dosage $d$ in figure 2. This is a purely frequentist method of evaluating the chances of survival.

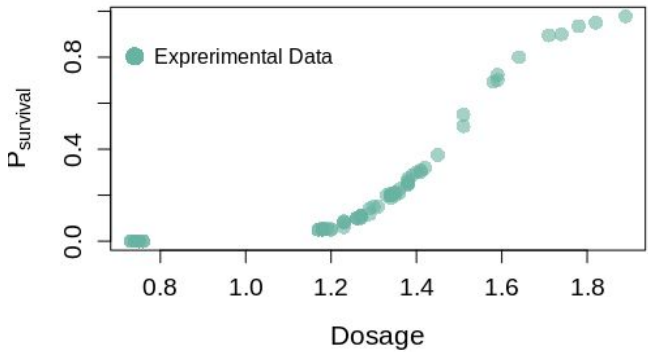

Fig. 2. Survival Ratio vs. Dosage

There are several flaws to this approach. First, for the results of a frequentist method to be somewhat reliable, a lot more data is needed, which means more experiments need to be conducted. Second, no sound estimation can be made for the chances of survival corresponding to out of range dosages, as there is no data in those ranges.

*2.3. Models*

There are different approaches that can be taken to model the problem. First, a dose-response model is proposed using the Hill equation, which is vastly used in biochemistry and pharmacology. Then, a simple logistic regression model, and a hierarchical logistic regression model are presented in a probabilistic framework using MCMC. All models are implemented in R (All codes can be found via the github repository at https://github.com/Dorsa-Arezooji/Sick-Pigs .)

*2.3.1 Dose-Response Model*

Many dose-response models use the Hill equation to describe nonlinear dose-response relationships (Goutelle *et al.*, 2008). These relationships are conveniently characterized by (generally sigmoidal) curves called dose-response curves that represent the response of the receptor in the y axis and the logarithm of the dosage in the x axis. These curves can be fit to the Hill equation, which is a logistic function similar to the logit function (Magnus, 1957). According to the Hill equation, the magnitude of response $R$ can be calculated for each dosage $d$, using the following:

$$E(d) = \frac{R_{max}}{1 + (\frac{d_{50}}{d})^{c_H}}$$

where $R_{max}$ denotes the maximum response, i.e highest survival ratio, $d_{50}$ denotes the dosage that corresponds to a 50% maximal response, and $c_H$ is the Hill coefficient calculated by the following formula:

$$c_H = \frac{\log(81)}{\log(\frac{d_{90}}{d_{10}})}$$

The resulting dose-response curve after applying the Hill equation to the dataset, illustrated in figure 3, shows a close fit to the data. Although there is a probabilistic interpretation of the Hill equation (Goutelle *et al.*, 2008), it does not have any direct Bayesian interpretation. Further, the accuracy of the fit is not ideal in higher dosages. Hence it can be improved using the Bayesian approach with MCMC sampling, as it will be demonstrated in the following sections.

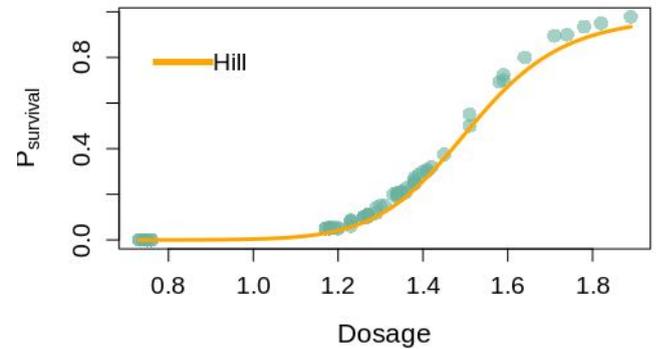

Fig. 3. Dose-Response Curve

*2.3.2 Simple Logistic Regression Model*

A Bayesian logistic regression model is proposed and implemented in R using the rstan package. The model takes parameters $\alpha$ and $\beta$, and dosage $d$ to construct the linear transformation $\alpha + \beta d$. It then applies the inverse logistic function $logit^{-1}$ to the linear transformation of the dosage and calculates the binomial likelihood of the data given the parameters. The Bayesian network (BN) of the problem is illustrated in figure 4. The dosage, number of subjects, and the number of improved subjects for an experiment $i$ are denoted by $d_i, N_i, n_i$ respectively. The parameters of the model are $\alpha$ and $\beta$, and $E$ is the total number of experiments. $\gamma_i$ denotes the probability of survival for a dosage $d_i$.

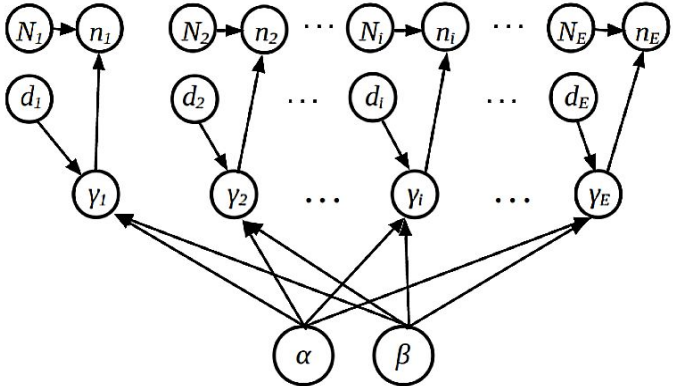

Fig.4.    Simple Logistic Regression BN

Assuming normal priors for convenience:

$$p(\alpha) = \mathcal{N}(\mu_\alpha, \sigma_\alpha) \quad p(\beta) = \mathcal{N}(\mu_\beta, \sigma_\beta)$$

and survival probability density $p_{survival}$ and dosage $d$

$$logit(p_{survival}) = \log(\frac{p_{survival}}{1 - p_{survival}}) = \alpha + \beta d$$

$$\gamma = p_{survival} = logit^{-1}(\alpha + \beta d) = \frac{1}{1 + e^{-(\alpha + \beta d)}}$$

the binomial likelihood for each node is calculated by

$$p(n, N, d | \alpha, \beta) = \binom{N}{n} \left(\frac{1}{1 + e^{-(\alpha + \beta d)}}\right)^n \left(\frac{e^{-(\alpha + \beta d)}}{1 + e^{-(\alpha + \beta d)}}\right)^N$$

and the model's likelihood being the product of all node likelihoods, the posterior is evaluated by

$$p(\alpha, \beta | n, N, d) = \frac{p(n, N, d | \alpha, \beta) p(\alpha) p(\beta)}{\iint_{a,b} p(n, N, d | a, b) p(a) p(b) \, da \, db}$$

After defining the data $d, N, n$, parameters $\alpha, \beta$, and the model's equations (priors and likelihood), the stan() function from rstan is evoked to generate samples from the posterior using HMC with 4 chains. By default, stan() takes half of total iterations for warm-up, then it will proceed with the sampling using the other half. Stan generates sampled output values for all of the specified parameters. For the results to be reliable, the chains must have mixed and converged ( $Rhat \leqslant 1$).

To study the effects of the prior, a number of different priors (normal, logistic, uniform) have been incorporated into the model (for stan codes see appendix A). The posteriors are generally similar for different choices of priors, including non-informative or "flat" priors. This is illustrated in figure 5. For more details about other priors see table 1 in appendix C.

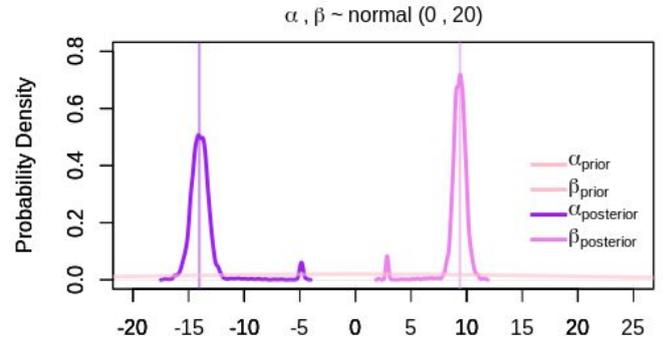

Fig.5.    Non-informative Priors - Simple LR Model

*2.3.3 Hierarchical Logistic Regression Model*

The same approach as the simple logistic regression is taken for the hierarchical mode. The BN of the model is shown in figure 6. In contrast with the previous model, now each group corresponds with a set of local $\alpha$ and $\beta$ which are updated by a set of global hyperpriors $\mu_\alpha, \sigma_\alpha$ and $\mu_\beta, \sigma_\beta$. The stan code can be found in appendix A.

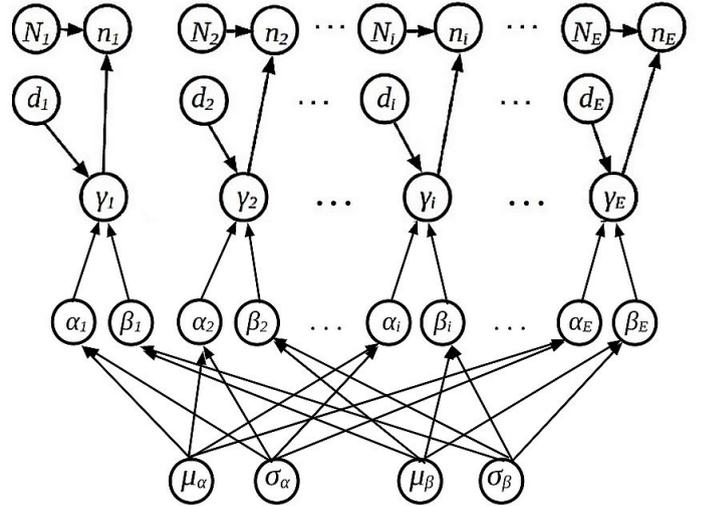

Fig.6.    Hierarchical Logistic Regression BN

As it was pointed out earlier, NCP is implemented to break correlations into deterministic transformations. In this model, NCP is crucial as without it, the Markov chains would not converge and sampling would get stuck in what is known as Neal's funnel (Neal, 2003). The global mean values for $\alpha$ and $\beta$ (-14.03 and 9.39 respectively), are quite close to the ones from the simple logistic regression model. The global posterior densities for $\mu_\alpha$ and $\mu_\beta$ are plotted in figure 7.

Figure 8 shows a comparison between the Hill model and the logistic regression model. It can be deduced that the Bayesian approach can be used to more accurately model the problem and continuously learn from the data to update posterior beliefs.

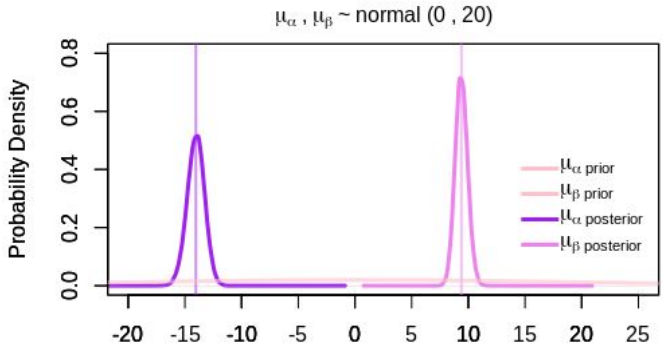

Fig.7. Non-informative Priors - Hierarchical LR Model

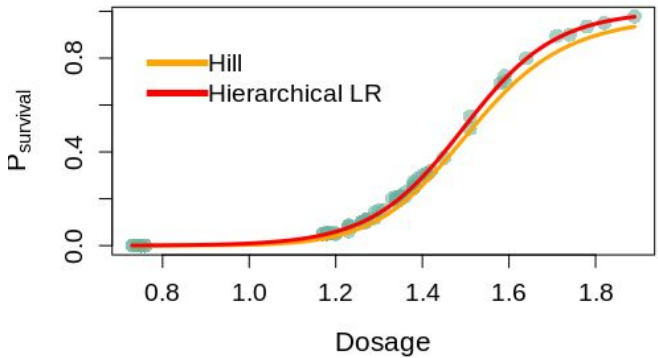

Fig.8. Comparison of Models in RStan

*2.4. Comparison with PyMC and AgenaRisk*

Both logistic regression models (hierarchical and simple) have been implemented in PyMC (Salvatier, Wiecki and Fonnesbeck, 2016), which also uses MCMC, and AgenaRisk (Agena ltd., 2018), which uses dynamic discretization.

In the dynamic discretization algorithm proposed by Kozlov and Koller (1997), the ranges of continuous nodes in the BN are searched to find the most accurate specification of the zones with the highest probability densities, also referred to as high density regions (HDR), based on the model and evidence. Then, in each iteration, a sequence of discretization intervals are obtained in the range of possible values for each corresponding node. After each iteration, the resulting discretization is evaluated to test if discretized probability densities converge to their true densities according to an acceptance criteria. Compared to static discretization, dynamic discretization of Bayesian models helps avoid unnecessary computations for the low density regions, thus saving storage and speeding up the process (Neil et. al, 2007). Additionally, the discretization can be adjusted once new evidence enters the BN, yielding a better accuracy than static discretization (Neil et. al, 2007).

Neil et al. (2007), modeled a hierarchical logistic regression BN using dynamic discretization. They also compared the performance of AgenaRisk and WinBUGS, concluding that unlike dynamic discretization, WinBUGS crashes in case of using non-informative priors. However, the comparison between AgenaRisk, RStan, and PyMC (table 1 appendix C) shows that for the simple logistic regression model with different priors, sampling with MCMC takes under a minute, while it takes several minutes with dynamic discretization. It should be noted that this comparison might not be accurate as the models were implemented in different machines with different hardware specifications.

The table below summarizes the hierarchical models's posterior results for RStan and AgenaRisk with priors $\mu_{\alpha,\beta} = normal(0, 20)$. Interestingly, RStan performs approximately 7x faster, and provides more accurate results than Agenarisk for the hierarchical model.

Table 2. Hierarchical Model in RStan and AgenaRisk

| Environment | *RStan* | | *AgenaRisk* | |
|---|---|---|---|---|
| **Runtime** | 4.6 minutes | | 34 minutes | |
| | $\mu\alpha$ | $\mu\beta$ | $\mu\alpha$ | $\mu\beta$ |
| **1st Qu.** | -14.53 | 9.02 | -15.71 | 10.35 |
| **Mean** | -14.01 | 9.38 | -15.55 | 10.48 |
| **3rd Qu.** | -13.51 | 9.75 | -15.36 | 10.56 |
| **sd** | 0.76 | 0.54 | 0.25 | 0.18 |
| | $\sigma\alpha$ | $\sigma\beta$ | $\sigma\alpha$ | $\sigma\beta$ |
| **1st Qu.** | 0.02 | 0.02 | 0.006 | 0.003 |
| **Mean** | 0.06 | 0.04 | 0.02 | 0.01 |
| **3rd Qu.** | 0.09 | 0.06 | 0.03 | 0.01 |
| **sd** | 0.05 | 0.03 | 0.03 | 0.01 |

Figure 9 shows the posterior densities of $\mu_\alpha$ and $\mu_\beta$ for the aforementioned hierarchical model. It can be deduced that while RStan evaluates the posteriors more accurately, the densities are flatter (with lower certainty) than AgenaRisk. The lack of optimal accuracy in AgenaRisk could be due to the fact that the simulation convergence acceptance criteria had been increased to 0.1 from the default value (0.001) to obtain results in a reasonable amount of time.

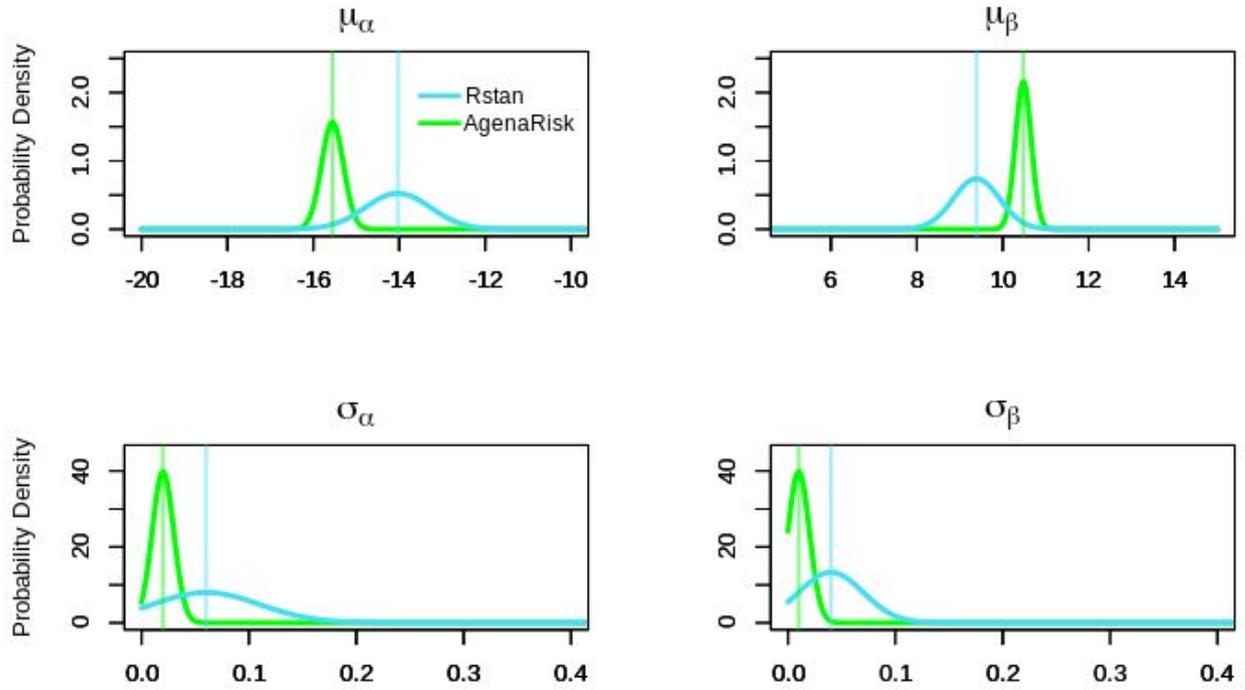

Fig.9. Comparison of Global Hyperparameters in RStan and AgenaRisk

## 3. Discussion

One of the assumptions in modeling the problem was that the probability of survival corresponding to each dosage can be represented by a binomial distribution. It should be noted that the binomial distribution assumes that independent samples are drawn "with replacement". If a number of samples $N_S$ are drawn from a population of size $N_P$ without replacement, the assumption of independence no longer holds true and the problem should be modelled by a hypergeometric distribution (Shuster, 2005). In the case where $N_S << N_P$, the binomial distribution is still an adequate approximation, hence it is widely used. The use of a binomial probability distribution in this paper implies that each pig may have been a part of more than one experiment, or each pig has only been tested on once but there is a substantially large population of pigs from which the subjects of this experiment have been picked out.

Another point of interest is that the comparison of certain performance metrics, specially computation time, are not reliable as the simulations have been carried out in different machines. However, it has been attempted to mitigate this issue for RStan and PyMC by assigning 4 CPU cores to computations in both environments. Furthermore, for the comparisons to be unbiased, the acceptance criteria for determining convergence needs to be similar. In the case that it is not, the results are prone to error.

## 4. Conclusion and Future Work

In this paper, a logistic regression model has been proposed and implemented in a probabilistic framework to model the effects of a drug. The problem has been modeled using two different approaches: MCMC sampling, and dynamic discretization. The proposed model's potential to fit the data has been compared with that of a widely used model based on the Hill equation which is conveniently used for dose-response studies. The results show enhanced accuracy of fit in the proposed Bayesian model. Further, a hierarchical logistic regression model was introduced, the results of which were on par with the non-hierarchical model. R

The comparative study between the implementation of MCMC in RStan and PyMC with that of dynamic discretization in AgenaRisk shows higher accuracy and efficiency (computation and storage) for hierarchical modeling with MCMC. Nevertheless, AgenaRisk can be used for a wider selection of priors compared to RStan and PyMC. For instance, both beta and Weibull priors can be chosen in AgenaRisk as it allows defining a custom range for the target distribution, [0, 1] for beta and [0, +inf] for Weibull.

The work in this paper can be expanded to model more complex problems in drug trials. For example, instead of binary outcomes (survived, not survived or 0,1), more expressive outcomes can be introduced to the model, such as the time it takes for each subject to improve to a certain extent. It is clear that such complex models come with more hurdles and considerations.

## 5. Acknowledgements

I would like to thank my supervisor, professor Martin Neil for continuously guiding me through this project. I would also like to thank Aadam Qaiser Khan and Minh Duc Phan for their contributions to this project by providing their results from AgenaRisk and PyMC

## 6. Appendix

All of the stan and R codes used for modeling, analysis, and visualization, along with their outputs are publicly available at the [github repo](). Alternatively, please read the [R notebook]() with all of the codes and results.

*Appendix A: codes*

```
# Non-Hierarchical Logistic Regression

m_LR = '
data {
  int<lower=1> E;
  vector[E] x;
  int<lower=1> N[E];
  int<lower=0> n [E];
}
parameters {
  real alpha;
  real beta;
}
model {
  alpha ~ dist_a;
  beta ~ dist_b;
  n ~ binomial_logit(N, alpha + beta * x);
}
'
```

```
# Hierarchical Logistic Regression without NCP

m_HLR = '
data {
  int<lower=1> E;
  vector[E] x;
  int<lower=1> N[E];
  int<lower=0> n[E];
}
parameters {
  vector[E] alpha;
  vector[E] beta;
  real mu_a;
  real mu_b;
  real<lower=0> sigma_a;
  real<lower=0> sigma_b;
}
model {
  mu_a ~ normal(0,20);
  mu_b ~ normal(0,20);
  sigma_a ~ normal(0,2);
  sigma_b ~ normal(0,2);
  alpha ~ normal(mu_a, sigma_a);
  beta ~ normal(mu_b, sigma_b);
  n ~ binomial_logit(N, alpha + beta .* x);
}
'
```

```
# Hierarchical Logistic Regression with NCP

m_H = '
data {
  int<lower=1> E;
  vector[E] x;
  int<lower=1> N[E];
  int<lower=0> n[E];
}
parameters {
  vector[E] a_raw;
  vector[E] b_raw;
  real mu_a;
  real mu_b;
  real<lower=0> sigma_a;
  real<lower=0> sigma_b;
}
transformed parameters {
  vector[E] alpha = mu_a + sigma_a * a_raw;
  vector[E] beta = mu_b + sigma_b * b_raw;
}
model {
  mu_a ~ normal(0,20);
  mu_b ~ normal(0,20);
  sigma_a ~ normal(0,2);
  sigma_b ~ normal(0,2);
  a_raw ~ std_normal();
  b_raw ~ std_normal();
  n ~ binomial_logit(N, alpha + beta .* x);
}
'
```

*Appendix B: plots*

A few non-hierarchical priors (RStan) and their posterior densities for reference. For more priors see the [github repository](#).

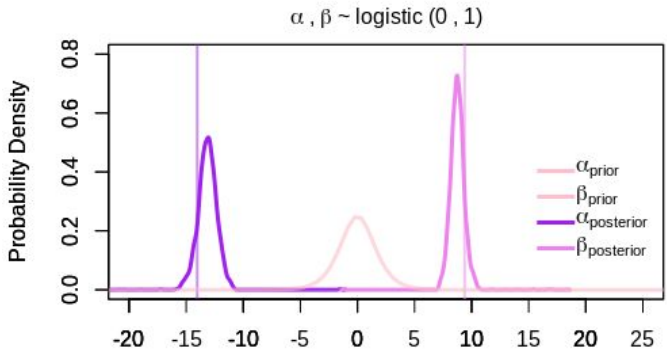

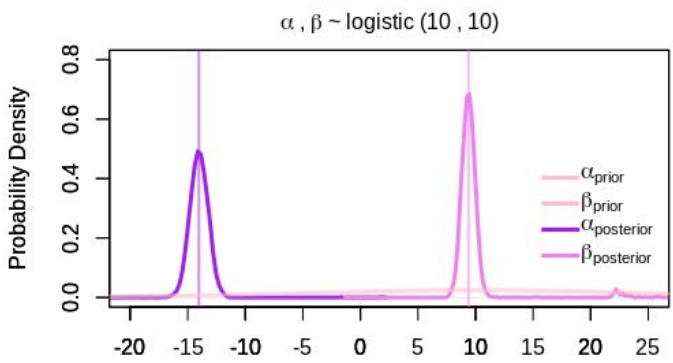

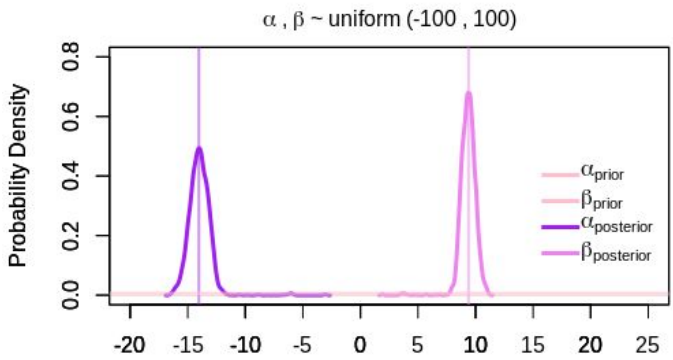

A few traceplots for reference to demonstrate that the chains have indeed mixed, hence the results are reliable (i.e. $Rhat \leqslant 1$).

i. A non-hierarchical model (m5 in the code):

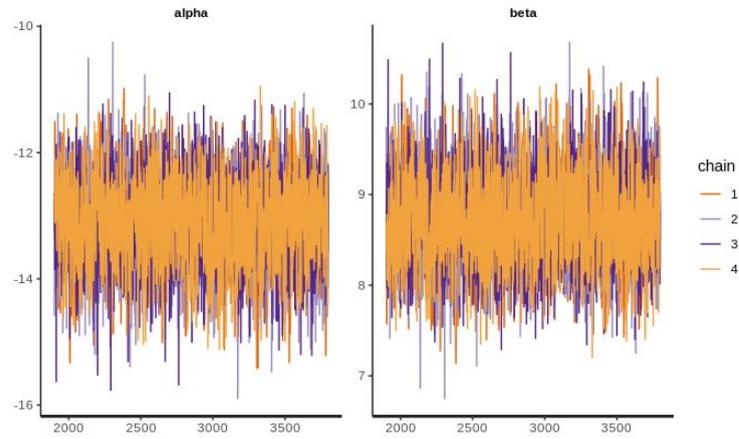

ii. A hierarchical model (m_hh in the code):

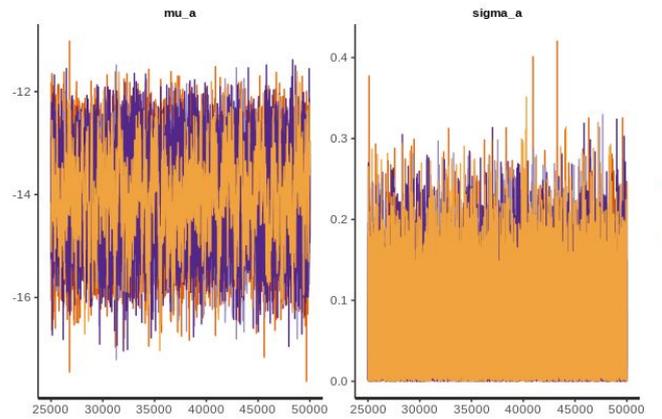

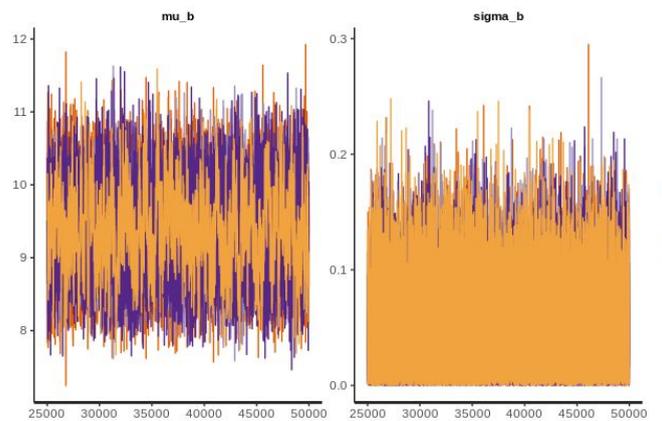

*Appendix C: tables*

Table 1. Comparison of Non-Hierarchical Result

| Prior | Iterations | | | Time Taken (mins:seconds) | | | $\alpha_{posterior}$ | | | $\beta_{posterior}$ | | |
|---|---|---|---|---|---|---|---|---|---|---|---|---|
| $\alpha, \beta$ | RStan | PyMC | AgenaRisk | RStan | PyMC | AgenaRisk | RStan | PyMC | AgenaRisk | RStan | PyMC | AgenaRisk |
| Beta (0.5, 0.5, -100, 100) | - | - | 29 | - | - | 29:00 | - | - | -14.17 | - | - | 9.50 |
| Beta (0.5, 0.5, -20, 20) | - | - | 16 | - | - | 11:00 | - | - | -14.27 | - | - | 9.56 |
| Beta (1, 1, -20, 20) | - | - | 17 | - | - | 11:11 | - | - | -14.16 | - | - | 9.49 |
| Beta (100, 100, -20, 20) | - | - | 14 | - | - | 13:25 | - | - | -10.10 | - | - | 6.58 |
| Logistic (0, 1) | 3800 | 13000 | 24 | 00:10 | 01:19 | 21:08 | -13.13 | -13.48 | -12.82 | 8.74 | 8.99 | 8.52 |
| Logistic (0, 10) | 3800 | 9000 | 23 | 00:14 | 00:48 | 19:00 | -13.97 | -14.01 | -14.11 | 9.35 | 9.38 | 9.45 |
| Logistic (10, 10) | 3800 | 8000 | 32 | 00:13 | 00:46 | 34:00 | -14.00 | -14.02 | -14.14 | 9.37 | 9.38 | 9.47 |
| Normal (0, 20) | 4000 | 6000 | 23 | 00:14 | 00:33 | 19:00 | -14.01 | -14.03 | -13.46 | 9.38 | 9.39 | 8.98 |
| Normal (0, 1) | 4200 | 13000 | - | 00:11 | 00:57 | - | -9.04 | -9.92 | - | 5.83 | 6.47 | - |
| Normal (0, 100) | 4000 | 5000 | 21 | 00:16 | 00:34 | 16:12 | -14.04 | -14.03 | -14.10 | 9.40 | 9.39 | 9.44 |
| Normal (0, 10000) | 4200 | 6000 | 45 | 00:17 | 00:40 | 45:27 | -14.04 | -14.03 | -14.16 | 9.40 | 9.39 | 9.48 |
| Normal (-100, 100) | 3850 | - | - | 00:16 | - | - | -14.03 | - | - | 9.39 | - | - |
| Uniform (-100, 100) | 3850 | - | 21 | 00:13 | - | 15:00 | -14.03 | - | -14.16 | 9.39 | - | 9.49 |
| Uniform (-1000, 1000) | 4000 | - | 45 | 00:14 | - | 48:21 | -14.05 | - | -14.16 | 9.41 | - | 9.49 |
| Uniform (0, 10) | - | - | 94 | - | - | 23:10 | - | - | 0.00 | - | - | 0.00 |
| Uniform (0, 20) | - | - | - | - | - | - | - | - | - | - | - | - |
| Uniform (0, 50) | - | - | - | - | - | - | - | - | - | - | - | - |
| Uniform (0, 100) | - | - | 83 | - | - | 25:00 | - | - | 0.00 | - | - | 0.00 |
| Uniform (0, inf) | - | - | - | - | - | - | - | - | - | - | - | - |
| Uniform (-inf, inf) | 3800 | 6000 | - | 00:19 | 00:37 | - | -14.00 | -14.05 | - | 9.37 | 9.41 | - |
| Weibull (1, 1) | - | - | 101 | - | - | 25:30 | - | - | 0.00 | - | - | 0.00 |